\documentstyle[epsfig,aps,pra,preprint]{revtex}

\begin{document}

\draft
\title{Predictability of large future changes in a \\ 
competitive evolving population
}
\author{D. Lamper and S. Howison}
\address{Oxford Centre for Industrial and Applied Mathematics, \\
Oxford University, Oxford, OX1 3LB, U.K.}
\author{N. F. Johnson}
\address{Physics Department, Oxford University, Oxford, OX1 3PU, U.K.}

\date{\today}

\maketitle

\begin{abstract}
The dynamical evolution of many economic, sociological, biological and
physical systems tends to be dominated by a relatively small number of
unexpected, large changes (`extreme events'). We study the large,
internal changes produced in a generic multi-agent population competing
for a limited resource, and find that the level of predictability
actually \emph{increases} prior to a large change. These large
changes hence arise as a predictable
consequence of information encoded in the system's global state.

\end{abstract}

\pacs{01.75.+m, 02.50.Le, 05.40.+j, 87.23.Ge
}


\newpage

Populations comprising many `agents' (e.g.\ people, species,
data-packets, cells) who compete for a limited resource, are
believed to underlie the complex dynamics observed in areas as
diverse as economics~\cite{lux99,econo,mantegna00,bouchaud00},
sociology~\cite{arthur94}, internet traffic~\cite{huberman98},
ecology~\cite{tella00} and biology~\cite{nowak92,sornette99}. The
reliable prediction of large future changes (`extreme events') in
such complex systems would be of enormous practical
importance, but is widely considered to be
impossible~\cite{malkiel85}. 

In this paper, we examine the predictability of large future changes
produced within an evolving
population of agents who compete for a limited
resource. We find that the level of predictability in the system
actually 
\emph{increases} prior to a large change. The implication is
that such a large change arises as a predictable consequence of
information encoded in the system's global state, as opposed to being
triggered by some isolated random event.

We consider a generic multi-agent system comprising a population
of $N_{tot}$ agents where only a maximum of $L < N_{tot}$ agents
can be winners at each timestep; an everyday example would be a
popular bar with a limited seating capacity $L$~\cite{arthur94}. For the
purpose of this paper, we consider a specific case of such a
limited-resource problem with $L = (N_{tot} - 1)/2$ with $N_{tot}$
being odd~\cite{challet97}, hence there are more losers
than winners. We note that similar dynamics can also occur for more
general
$L$ values~\cite{johnson99}. Each agent is therefore seeking
to be in the minority group: for example, a buyer in a financial market
may obtain a better price if more people are selling than buying; a
driver may have a quicker journey if she chooses the route with less
traffic. At each timestep, an agent decides whether to enter a game where
the choices are option 0 (e.g.\ buy, choose route A) and option 1 (e.g.\
sell, choose route B). Each agent holds a finite number of strategies and
only a subset $N=N_{0}+N_{1}\leq N_{tot}$ of the population, who are
sufficiently confident of winning, actually play: $N_{0}$ agents
choose 0 while $N_{1}$ choose 1. If $N_{0}-N_{1}>0$, the winning
decision (outcome) is `1' and vice versa. If $N_{0}=N_{1}$ the
tie is decided by a coin-toss. Hence $N$ and the `excess demand'
$N_{0-1}=N_{0}-N_{1}$ both fluctuate with time. In contrast to
the basic Minority Game (MG)~\cite{challet97}, this variable-$N$
model has the realistic feature of accounting for agents'
confidence~\cite{johnson01,johnson00}. Furthermore the variable-$N$ model
can be used to generate statistical and dynamical features similar to
those observed in financial markets (archetypal
examples of complex systems)~\cite{johnson01,econo}.
Therefore, demonstration of predictability of extreme events in the
present multi-agent model would open up the exciting possibility of
predictability of extreme events in real-world systems.

The only global information available to the agents is a common
bit-string `memory' of the $m$ most recent outcomes. The agents can
thus be said to exhibit `bounded rationality'~\cite{arthur94}. Consider
$m=2$; the $2^{m}=4$ possible history bit-strings are $00,01,10$
and $11$. A strategy consists of a response, i.e.\ 0 or 1, to each
possible bit-string; hence there are  $2^{2^{m}}=16$ possible
strategies. At the beginning of the game, each agent randomly
picks $q$ strategies and after each turn assigns one (virtual)
point to a strategy which would have predicted the correct
outcome. Agents have a time horizon $T$ over which virtual points
are collected and a threshold probability level $\tau$;
strategies with a probability of winning greater than or equal to
$\tau$, i.e.\ having $\geq T \tau $ virtual points, are available to
be used by the agent. We call these \emph{active} strategies.
Agents with no active strategies within their individual set of
$q$ strategies do not play at that timestep. Agents with one or
more active strategies play the one with the highest virtual point
score; any ties between active strategies are resolved using a
coin-toss. The `excess demand' $N_{0-1}$, which can be identified
as the output from the model system, can be expressed as
\begin{eqnarray} \label{eq:N}
N_{0-1} &=& \sum_i \Big\{ 1-2 x_{i} s_{i} \Big\}
\end{eqnarray}
where $s_{i}$ is the prediction of the $i$-th strategy, e.g.\ 0 or
1, and $x_{i}$ is the number of agents using this strategy, with
the summation taken over the set of active strategies at that
timestep.

Because of the feedback in the game, any particular strategy's
success is short-lived. If all the agents begin to use similar
strategies, and hence make the same decision, such a strategy
ceases to be profitable.  The game can be broadly classified into
three regimes. (i) The number of strategies in play is much
greater than the total available: groups of traders will play
using the same strategy and therefore crowds should dominate the
game~\cite{johnson99a}. (ii) The number of strategies in play is
much less than the total available: grouping behaviour is
therefore minimal. (iii) The number of strategies in play is
comparable to the total number available: this represents a
transition regime and is of most interest, since it produces
seemingly random dynamics with occasional large movements. Even
if complete knowledge of the state of the game were available at
any timestep, it seems impossible that subsequent outcomes should
be predictable with significant accuracy since the coin-tosses
which are used to resolve ties in decisions (i.e.
$N_{0}$=$N_{1}$) and active-strategy scores inject stochasticity
into the game's time-evolution. Remarkably, however, we find that
\emph{large} changes over several consecutive timesteps can be
predicted with surprising accuracy \emph{without} any detailed
knowledge of the game itself.

Suppose we are given an analogue time series $H(t)$ generated by a
physical, sociological, biological or economic system, e.g.\ a
financial market~\cite{johnson01}, whose dynamics are well-described by the
multi-agent game for a fixed \emph{unknown} parameter set $m, N,
\tau, T$ and an \emph{unknown} specific realization of initial
strategy choices. We call this our `black-box' game. Our goal is
to identify `third-party' games which can be used to predict large
future changes in $H(t)$, where $\Delta H(t)$ is defined to be
directly proportional to the excess demand $N_{0-1}$.  For
example, $\Delta H(t)$ could be the price change in a financial
market, or may instead be a quantity which is derived from the
system output using a known non-linear function.  For the
remainder of this article, we focus on the following game
parameters for the `black-box' game: $N=101$, $m=3$, $q=2$,
$T=100$, $\tau=0.53$, although our conclusions are more
general~\cite{supplement}. Since $\tau > 0.5$ an agent will not
participate unless she believes she has a better than average chance of
winning.

\begin{figure}
\centering
\centerline{\epsfig{file=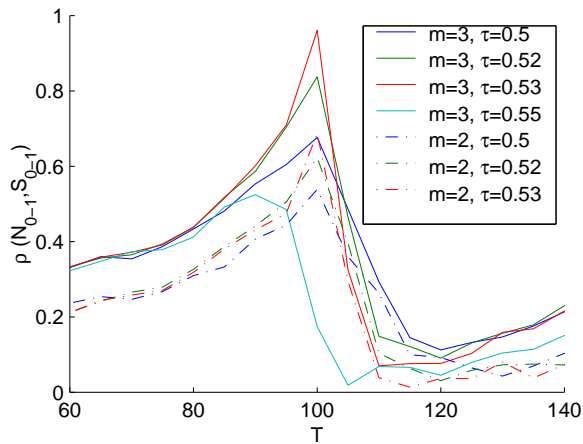,width=7.8cm}}
\vspace{0.2cm}
\caption{
  Estimation of the parameter set for the black-box game.  The
  correlation between $N_{0-1}$ and $S_{0-1}$ is calculated over 200 timesteps
  for an ensemble of candidate third-party games.  The third-party
  game that achieves the highest correlation is the one with the same
  parameters as the black-box game.}
\label{fig1}
\end{figure}

We start by running $H(t)$ through a trial third-party game in
order to generate an estimate of $S_{0}$ and $S_{1}$ at each
timestep, the number of active strategies predicting a 0 or 1
respectively. This is obtained from the strategy space, or the
pool of all available strategies in the third party game, and is
independent of the distribution of agents. We wish to predict
$\Delta H(t)$, i.e.\ $N_{0-1}$ ; we will do this by linking $S$
and $N$ through an appropriate probability distribution. Provided
the strategy space in the black-box game is reasonably well
covered by the agent's random choice of initial strategies, any
bias towards a particular outcome in the active strategy set will
propagate itself as a bias in the value of $N_{0-1}$ away from
zero. Thus we expect $N_{0-1}$ to be approximately proportional
to $S_{0}-S_{1}=S_{0-1}$.  This is equivalent to assuming an
equal weighting $x_{i}$ on each strategy in Eq.\ (\ref{eq:N}),
indicating the exact distribution of strategies among the
individual agents is unimportant in this regime~\cite{kalman}. In addition, the number of agents taking part in the game at each
timestep will be related to the total number of active strategies
$S_{0}+S_{1}=S_{0+1}$, hence the error (i.e.\ variance) in the
prediction of $N_{0-1}$ using $S_{0-1}$ will be dependent on
$S_{0+1}$. Based on extensive statistical analysis of known
simulations for the multi-agent game~\cite{supplement}, we have confirmed
that it is reasonable to model the relationship by
\begin{eqnarray*}
N_{0-1}&=&b S_{0-1} + \varepsilon[0,f(S_{0+1})]
\end{eqnarray*}
where $\varepsilon$ is a noise term with mean zero and variance
proportional to $S_{0+1}$, and $b$ is a constant. In particular,
$N_{0-1}$ is well described by a Normal distribution of the form
$N_{0-1} \sim \mbox{N}(b S_{0-1}, c S_{0+1})$, where $c$ is a
constant. The variance of our forecast density function can be
minimized by choosing a third-party game that achieves the
maximum correlation between $N_{0-1}$ and our explanatory
variable $S_{0-1}$, with the unexplained variance being
characterized by a linear function of $S_{0+1}$. We focus on the
parameter regime known to produce realistic statistics (e.g. 
fat-tailed distribution of returns in financial markets).  Within this
parameter space we run an ensemble of third-party games through the
black-box series
$H(t)$, calculating the values of
$S_{0-1}$ from the reconstructed strategy space. We then identify
the configuration that achieves the highest correlation between
$S_{0-1}$ and $N_{0-1}$ produced by the original black-box game.
As shown in Fig.\ \ref{fig1}, the third-party game that achieves the
highest correlation is the one whose parameters coincide with the
black-box game. From a knowledge of just $H(t)$, and hence
$N_{0-1}$, we have therefore used next-step prediction to recover
all the parameters of relevance to produce a `model' game for
prediction purposes.

We now extend this forecast to an arbitrary number $j$ of
timesteps into the future, in order to address the predictability
of large changes in $H(t)$ arising over several consecutive
timesteps. This is achieved by calculating the net value of
$S_{0-1}$ along all the $k=2^{j-1}$ possible future routes of the
third party game, weighted by appropriate probabilities. In order
to assign these probabilities, it is necessary to calculate all
possible $S_{0-1}$ values in the next $j$ timesteps. This is
possible since the only data required to update the strategy
space between timesteps is knowledge of the winning decision, and
hence the third party game can be directed along a given path
independent of the predictions of the individual agents in the
black box game. The change in $N_{0-1}$ along a path indexed by
$k$ is given by a convolution of the predictions over the $j$
individual steps
\begin{eqnarray*}
\mbox{N}(\mu_{k},\sigma_{k})&\sim&\mbox{N}\left(b\sum
S_{0-1}, c \sum S_{0+1} \right),
\end{eqnarray*}
where the summation is taken along the path represented by $k$.
In general, the pdf for the change in $N_{0-1}$ during the next
$j$ timesteps is a mixture of Normals:
\begin{eqnarray} \label{eq:prob}
\mbox{P}[\Delta
N_{0-1}(i;i+j)]&=&\sum_{k=0}^{2^{j}-1}p_{k}\mbox{N}(\mu_{k},\sigma_{k}),
\end{eqnarray}
where $p_{k}$ is the probability of path $k$ being taken.

To test the validity of the density forecast, we perform a
statistical evaluation using the realized variables.  The
one-step-ahead forecasts are normal distributions, and we define
the test statistic $Z_{i}$ as
\begin{eqnarray} \label{eq:Z}
Z_{i}&=&\frac{x_{i}-\mu_{i}^{x}}{\sigma_{i}^{x}}
\end{eqnarray}
where $\mu_{i}^{x}$ and $\sigma_{i}^{x}$ are the mean and variance of
the forecast distribution, and $x_{i}$ is the realised value of
$N_{0-1}$ at the timestep $i$. The $Z_{i}$ were found to be
independent uniform $N(0,1)$ variates for 1000 out-of-sample
predictions, confirming that the predicted distributions are correct.
To compare the forecasts to a naive `no-change' prediction, we
calculate the Theil coefficient~\cite{hellstrom} which is the sum of
squared prediction errors divided by the sum of squared errors
resulting from the naive forecast.  A coefficient of less than one
implies a superior performance compared to the naive prediction;
calculated values were typically in the region of 0.4. There is no
accepted method in the literature for evaluating multi-step-ahead
forecasts~\cite{tay00}. However, the density function for an arbitrary
time horizon is a mixture of Normal distributions, see Eq.\ (\ref{eq:prob}), 
each of which can be roughly characterised in terms
of a single mean and variance:
\begin{eqnarray*}
\mbox{E}[X]&=&\sum_{i=1}^{n}p_{i}\mu_{i} \\
\mbox{Var}[X]&=&\sum_{i=1}^{n}p_{i}(\sigma_{i}^{2}+\mu_{i}^{2})-\left(\sum_{i=1}^{n}p_{i}\mu_{i}\right)^{2}
\end{eqnarray*}
Hence the same test statistic as Eq.\ (\ref{eq:Z}) can be
calculated. Again, the predictions were found to be reliable.

\begin{figure}
\centering
\centerline{\epsfig{file=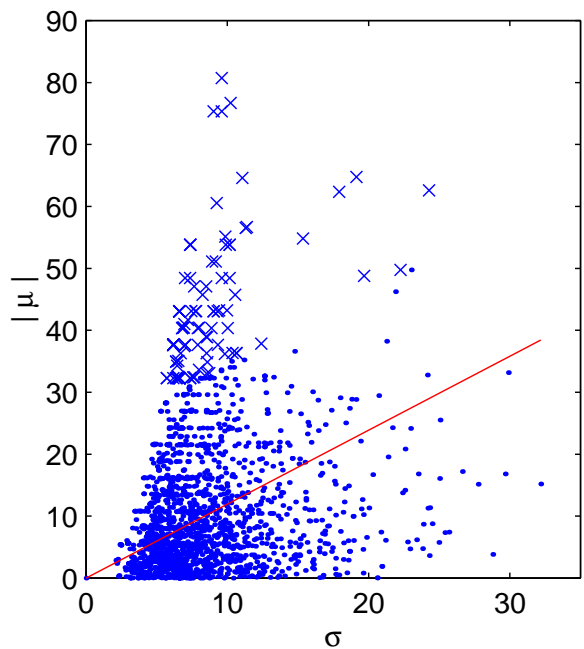,width=7.8cm}}
\vspace{0.2cm}
\caption{
  A plot of $\vert \mu \vert$ vs. $\sigma$ for 500 separate four-step density forecasts.
  Items marked by ``x'' are forecasts with an unusually large value of
  $\vert\mu\vert / \sigma$.  At these moments, the game is likely to be in a highly
  predictable phase.}
\label{fig2}
\end{figure}

\begin{figure}
\centering
\centerline{\epsfig{file=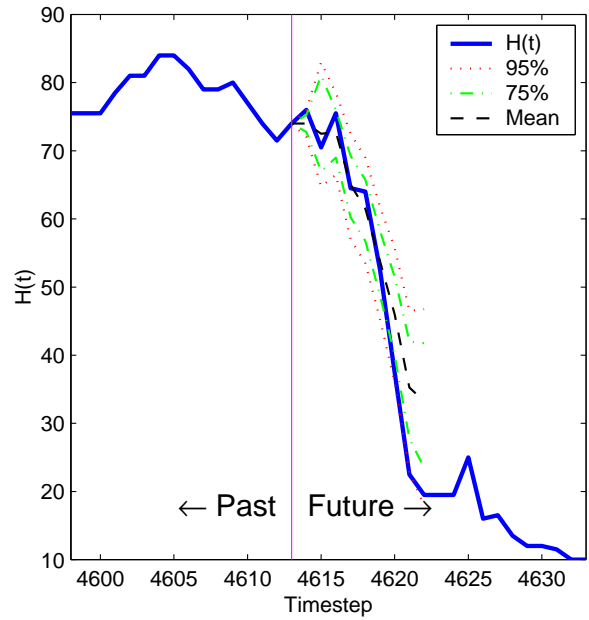,width=7.8cm}}
\vspace{0.2cm}
\caption{
 Comparison between the forecast density function and the
realised time series
  $H(t)$ for a typical large movement.  The large, well-defined
movement is
  correctly predicted.}
\label{fig3}
\end{figure}

Given that we can derive accurate distributions for the future changes
in $H(t)$, these will be of most practical interest in situations
where there is likely to be a substantial, well-defined movement.  We
characterise these moments by seeking distributions with a high value
of $\vert \mu \vert$ and a low value of $\sigma$ at a future timestep,
or over a specified time horizon. In Fig.\ \ref{fig2} we plot $\vert
\mu \vert$ vs.\ $\sigma$ for a number of separate forecasts, and take
a fraction of points that are furthest from the average trend
indicated by the regression line, i.e.\ we are interested in the
outliers.  The point with the highest residual is thus a candidate for
the game to be in a highly predictable phase. We call these time
periods \emph{predictable corridors}, since comparatively tight
confidence intervals can be drawn for the future evolution of the
excess demand, a typical example of which is shown in Fig.\ \ref{fig3}. We
subject these points to an identical test as described earlier to ensure
these potential outliers are well described by our probability
distributions, and this is found to be true.

We performed extensive numerical simulations to check the validity of
these predictive corridors~\cite{supplement}.  Our procedure is to take a
sample of 5000 timesteps, then fit parameters using the first 3000 steps.
We then look at the largest changes (extreme events) in our out-of-sample
region.  Extreme events are ranked by the largest movements in $H(t)$
over a given window size $W$. Hence we consider the top twenty extreme
events and calculate the probability integral transform $z_{t}$ of the
realized variables with respect to the forecast densities. The $z_{t}$
are found to be approximately uniform U[0,1] variates, confirming that
the forecast distribution is essentially correct - see supplementary
material for full details.  About 50\% of large movements occur in
periods with tight predictable corridors, i.e. a large value of
$\vert\mu\vert/\sigma$. Both the magnitude and sign of these extreme
events are therefore predictable. The remainder correspond to periods
with very wide corridors. Although the magnitude of the future
movement is now uncertain, the present method predicts with high
probability the actual direction of change. Even this more limited
information would be invaluable for assessing future risk in the
physical, economic, sociological or biological system of interest.
Our predictions generated from the third-party game were consistent
with all such extreme changes in the actual (black-box) time series
$H(t)$. Finally we note that some empirical support for our claim of
enhanced predictability prior to extreme movements, has very recently
appeared for the case of financial
markets~\cite{mansilla01}.

We are very grateful to Michael Hart, Paul Jefferies, Pak Ming
Hui and Jeff Dewynne for their involvement in this project. D.L.
thanks EPSRC for studentship.


\begin{thebibliography}{10}
\bibitem{supplement} Further details of the statistical tests and
parameter sets are available from authors.
\end{thebibliography}

\end{document}